\documentclass{aastex63}

\submitjournal{ApJ}

\begin{document}

\title{A $\Delta R\sim 9.5$ mag Super Flare of An Ultracool Star Detected by $\text{SVOM/GWAC}$ System}

\correspondingauthor{Jianyan Wei}
\email{wjy@nao.cas.cn}

\author{L. P. Xin}
\affiliation{CAS Key Laboratory of Space Astronomy and Technology, National Astronomical Observatories, Chinese Academy of Sciences, Beijing 100101, China.}

\author{H. L. Li}
\affiliation{CAS Key Laboratory of Space Astronomy and Technology, National Astronomical Observatories, Chinese Academy of Sciences, Beijing 100101, China.}

\author{J. Wang}
\affiliation{Guangxi Key Laboratory for Relativistic Astrophysics, School of Physical Science and Technology, Guangxi University, Nanning 530004, China}
\affiliation{CAS Key Laboratory of Space Astronomy and Technology, National Astronomical Observatories, Chinese Academy of Sciences, Beijing 100101, China.}

\author{X. H. Han}
\affiliation{CAS Key Laboratory of Space Astronomy and Technology, National Astronomical Observatories, Chinese Academy of Sciences, Beijing 100101, China.}

\author{Y. Xu}
\affiliation{CAS Key Laboratory of Space Astronomy and Technology, National Astronomical Observatories, Chinese Academy of Sciences, Beijing 100101, China.}

\author{X. M. Meng}
\affiliation{CAS Key Laboratory of Space Astronomy and Technology, National Astronomical Observatories, Chinese Academy of Sciences, Beijing 100101, China.}

\author{H. B. Cai}
\affiliation{CAS Key Laboratory of Space Astronomy and Technology, National Astronomical Observatories, Chinese Academy of Sciences, Beijing 100101, China.}

\author{ L. Huang}
\affiliation{CAS Key Laboratory of Space Astronomy and Technology, National Astronomical Observatories, Chinese Academy of Sciences, Beijing 100101, China.}

\author{X. M. Lu}
\affiliation{CAS Key Laboratory of Space Astronomy and Technology, National Astronomical Observatories, Chinese Academy of Sciences, Beijing 100101, China.}

\author{Y. L. Qiu}
\affiliation{CAS Key Laboratory of Space Astronomy and Technology, National Astronomical Observatories, Chinese Academy of Sciences, Beijing 100101, China.}


\author{X. G. Wang}
\affiliation{Guangxi Key Laboratory for Relativistic Astrophysics, School of Physical Science and Technology, Guangxi University, Nanning 530004, China}

\author{ E. W. Liang}
\affiliation{Guangxi Key Laboratory for Relativistic Astrophysics, School of Physical Science and Technology, Guangxi University, Nanning 530004, China}

\author{Z. G. Dai}
\affiliation{School of Astronomy and Space Science, Nanjing University, Nanjing 210093, China}
\affiliation{Key Laboratory of Modern Astronomy and Astrophysics (Nanjing University), Ministry of Education, Nanjing 210093, China}

\author{X. Y. Wang}
\affiliation{School of Astronomy and Space Science, Nanjing University, Nanjing 210093, China}
\affiliation{Key Laboratory of Modern Astronomy and Astrophysics (Nanjing University), Ministry of Education, Nanjing 210093, China}

\author{C. Wu}
\affiliation{CAS Key Laboratory of Space Astronomy and Technology, National Astronomical Observatories, Chinese Academy of Sciences, Beijing 100101, China.}

\author{J. B. Zhang}
\affiliation{Key Laboratory of Optical Astronomy, National Astronomical Observatories, Chinese Academy of Sciences, Beijing 100101, P.R. China}

\author{G. W. Li}
\affiliation{CAS Key Laboratory of Space Astronomy and Technology, National Astronomical Observatories, Chinese Academy of Sciences, Beijing 100101, China.}

\author{D. Turpin}
\affiliation{CAS Key Laboratory of Space Astronomy and Technology, National Astronomical Observatories, Chinese Academy of Sciences, Beijing 100101, China.}
\affiliation{Universit$\acute{e}$ Paris-Saclay, CNRS, CEA, D$\acute{e}$partement d'Astrophysique, Astrophysique, Instrumentation et Mod$\acute{e}$lisation de Paris-Saclay 91191, Gif-sur-Yvette, France.}

\author{Q. C. Feng}
\affiliation{CAS Key Laboratory of Space Astronomy and Technology, National Astronomical Observatories, Chinese Academy of Sciences, Beijing 100101, China.}

 \author{J. S. Deng}
 \affiliation{CAS Key Laboratory of Space Astronomy and Technology, National Astronomical Observatories, Chinese Academy of Sciences, Beijing 100101, China.}
 
 \affiliation{School of Astronomy and Space Science, University of Chinese Academy of Sciences, Beijing, China}

 \author{S. S. Sun}
 \affiliation{Guangxi Key Laboratory for Relativistic Astrophysics, School of Physical Science and Technology, Guangxi University, Nanning 530004, China}
\affiliation{CAS Key Laboratory of Space Astronomy and Technology, National Astronomical Observatories, Chinese Academy of Sciences, Beijing 100101, China.}
 \affiliation{School of Astronomy and Space Science, University of Chinese Academy of Sciences, Beijing, China}

  \author{ T. C. Zheng}
   \affiliation{Guangxi Key Laboratory for Relativistic Astrophysics, School of Physical Science and Technology, Guangxi University, Nanning 530004, China}

  \author{ Y. G. Yang}
  \affiliation{School of Physics and Electronic Information, Huaibei Normal University, Huaibei 235000, China. }

\author{ J. Y. Wei}
\affiliation{CAS Key Laboratory of Space Astronomy and Technology, National Astronomical Observatories, Chinese Academy of Sciences, Beijing 100101, China.}


\begin{abstract}

In this paper, we report the detection and follow-ups of a super stellar flare GWAC\,181229A with an amplitude of $\Delta R\sim$9.5 mag  on a M9 type star by $\text{SVOM/GWAC}$ and the dedicated follow-up telescopes. 
The estimated bolometric energy $E_{bol}$ is $(5.56-9.25)\times10^{34}$ ergs, which places the event to be one of the most powerful flares on ultracool stars. The magnetic strength is inferred to be (3.6-4.7) kG. 
Thanks to the sampling with a cadence of 15 seconds,   a new component near the peak time with a very steep decay is detected in the  $R$-band light curve, 
followed by the two-component flare template given by Davenport et al. (2014).   
An effective temperature of $5340\pm40$ K is measured by a blackbody shape fitting to the spectrum at the shallower phase during the flare.
The filling factors of the flare are estimated to be $\sim$30\% and 19\% at the peak time and  at 54 min after the first detection.
The detection of the particular event with large amplitude, huge-emitted energy and a new component demonstrates that a high cadence sky monitoring cooperating with fast follow-up observations is very essential for understanding the violent magnetic activity.

\end{abstract}

\keywords{flare --- stars: individual (GWAC\,181229A)---techniques: photometric--- techniques: spectroscopic}


\section{Introduction} \label{sec:intro}

The ultracool dwarfs (hereafter UCDs) are stars with spectral types later than M7 and mass below 0.3M$_\odot$. 
Empirically, UCDs are found to have weak chromospheric emission 
(Gizis et al.,  2000; Basri 2000) and be dim in the X-ray wavelength.
But the occurrence of flares on these stars at optical as well as X-ray (eg., Fleming et al. 2000), 
ultraviolet (eg., Linsky et al., 1995) and radio wavelengths show that  
magnetic activity does exist for very low-mass stellar configuration.
The interior of UCDs are presumably fully convective. It is proposed that the dynamo mechanisms 
for the chromospheric and coronal activity of these UCDs might be different from the solar-type stars ( Chabrier \& Baraffe 2000).

It is well known that the stellar flares are due to magnetic reconnection in a strong magnetic field (e.g, Shulyak et al., 2017).  However, 
during these stellar flares, the underly mechanism of the white light continuum is still not fully understood though lots of researches have been presented
including a  hydrogen recombination model (Kunkel  1969),
 a two-component model consisting hydrogen recombination and impulsively heated photosphere (Kunkel 1970), and 
a multi-component model (Zhilyaev et al.,  2007) in which
blackbody radiation are dominated at flare peak, and the hydrogen continuum are primarily during the flare decay. 
Gizis et al. (2013) proposed that the white-light emission mainly contributed by thermal continuum.

Thanks for the high cadence survey, like Kepler survey (Paudel et al. 2018) and ASAS-SNs (Schmidt et al. 2019),  
more late-type stellar 
flares were reported and analyzed in detail (Schmidt et al. 2019; Kowalski et al. 2010, 2013; Davenport 2016; Chang et al. 2018; Frith et al. 2013).
Paudel et al. (2018) pointed out that white-light flares are ubiquitous in M6-L0 dwarfs as seen in Kepler survey (Borucki et al. 2010) of ultracool dwarfs.
Schmidt et al. (2019) reported that the energy of M dwarf flares ranges from $10^{32}$ to $10^{35}$ erg after analyzing 47 ASAS-SN M dwarf flares. 
The occurrence rate of a flare with high energy ($E_U >10^{34}$) is expected once per month to year (Kowalski et al. 2010; Davenport et al. 2016;  
Rodriguez et al. 2018). 
These detections of flares of UCDs are helpful for understanding 
both the changes in the underlying magnetic dynamo and the interaction between the magnetic fields and surface of those ultracool stars.

Observationally, a white-light flare is typical of a rapid transient that is characterized by an initial impulsive rise with a duration of seconds and then 
by a decay with a timescale of seconds to hours (e.g., Davenport et al. 2014). 
Since the flares occur stochastically, an attractive method of detection is to monitor a large proportion of the sky by 
an automated survey with a cadence down to seconds. Ideally, the survey should have self-trigger capability and dedicated follow-up telescopes, which are 
required to capture the flares and to cover the total duration of the flares from the quiescent state before the start of the events to the time
at which the flares return back to the quiescent state.

In this paper, we report the detection of a super stellar flare with an amplitude of $\Delta R=9.5$ mag on a M9 star by GWAC system. Fast photometries and an optical spectrum for the flare were carried out. 
The total energy in $R$ band is about $E_R=1.5\times10^{34}$ erg.
This huge energy release places the event to be one of the strongest late-M dwarf flares up to now.
The paper is organized as follows. The discovery of the super flare is described in section 2. Section 3 
reports the rapid follow-ups by both photometry and spectroscopy. 
The properties of the flare are presented in section 4. Section 5 gives the discussion and summary for this discovery.

\section{Detection by GWAC}

\subsection{Detection and follow-up system of GWAC}
 As one of the main ground facilities of $\text{SVOM}$\footnote{SVOM is a China-France satellite mission dedicated to the detection and study of Gamma-ray bursts (GRBs)}
mission (Wei et al. 2016; Yu et al. 2020), GWAC (Ground-based Wide Angle Cameras) system located at Xinglong observatory of NAOC
is an optical transient survey  
that images the sky in optics down to $R\sim$16.0 mag at a cadence of 15 seconds, which aims to detect various of short-duration astronomical events, 
including the electromagnetic counterparts of gamma-ray bursts (Wei et al. 2016) and gravitational waves (Turpin et al. 2020), 
and stellar flares. The main characteristic and the survey strategy of GWAC is presented as follows. More detailed information of GWAC could be found in the reference (Wang et al., 2020).

The effective aperture size of each GWAC JFoV camera  is 18 cm. 
The f-ratio is $f/1.2$. Each camera is equipped with 4096$\times$4096 E2V 
back-illuminated CCD chip.
The wavelength range is from 0.5 to 0.85 $\mu m$.  
The field of view for each camera is 150 deg$^2$ and a pixel scale is 11.7 arc seconds.
For GWAC, each mount carries four JFoV cameras (an unit is called in GWAC system).
The total FoV for each unit is $\sim$ 600 deg$^2$.  
Currently, four units have been seted at Xinglong observatory, Chinese academy of Sciences, China. 
More units will be seted before the lunch of SVOM mission at 2022 aiming to cover about 5000 deg$^2$ simultaneously. 
During the survey, each  unit is assigned to a given grid which is pre-defined for the whole sky 
according to the FoV of each unit. The sky with a Galactic latitude of $ b < 20$  deg
as well as the grids
near the Moon are set with lower priority since the detection efficiency of any transient observing these sky 
will be reduced by the higher star density or higher background noise.

A dedicated rapid follow-up system has been developed for each candidate by using  two Guangxi-NAOC 60 cm optical telescopes (F60A and F60B) deployed beside GWAC 
with a typical delay time of one minute (Xu et al. 2020).
More deep imaging and spectroscopy can be carried out 
through Target of Opportunity observations by the 2.16 m telescope (Fan et al. 2016) at Xinglong observatory and by the
2.4m telescope at Gaomeigu observatory, China. 
The high cadence, middle detection limit, self-automatic trigger capability and its dedicated rapid follow-up telescopes enable GWAC system to 
detect a great number of stellar flares and to capture the events similar to  super flare ASASSN-16ae ( $\Delta V<11$ mag, Schmidt et al. 2016) 
with more intensive temporal resolution. 
 
 \subsection{Detection of the flare}
On 2018 December 29 UT10:42:51, an alert was generated by the GWAC on-line pipelines for a very bright optical transient (GWAC\,181229A)
during a survey for one pre-defined field from 10:03:07.8 to 14:55:21.0 UT at the same night.

\begin{figure}[htbp]
 \centering
 \includegraphics[width=0.3\textwidth]{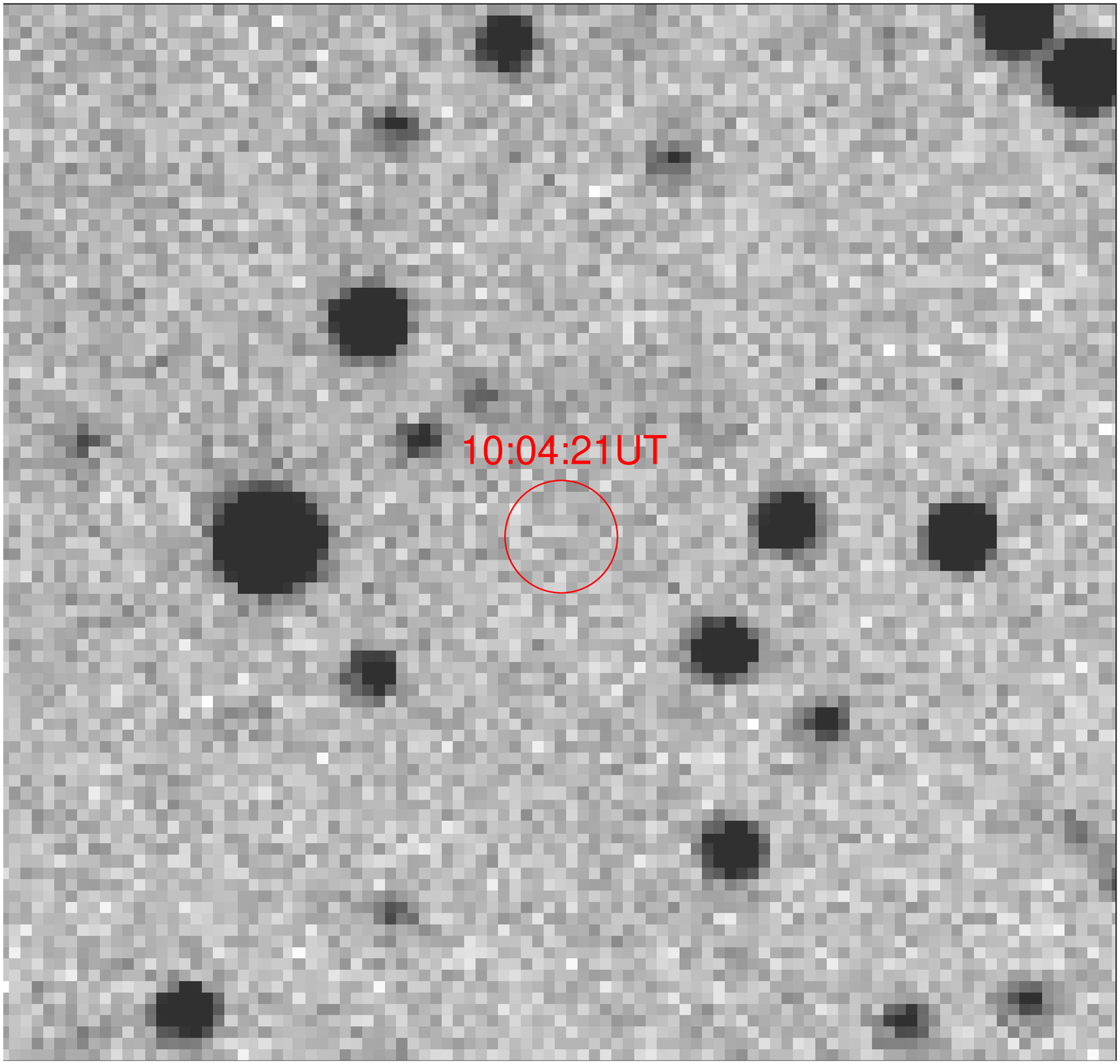}
  \includegraphics[width=0.3\textwidth]{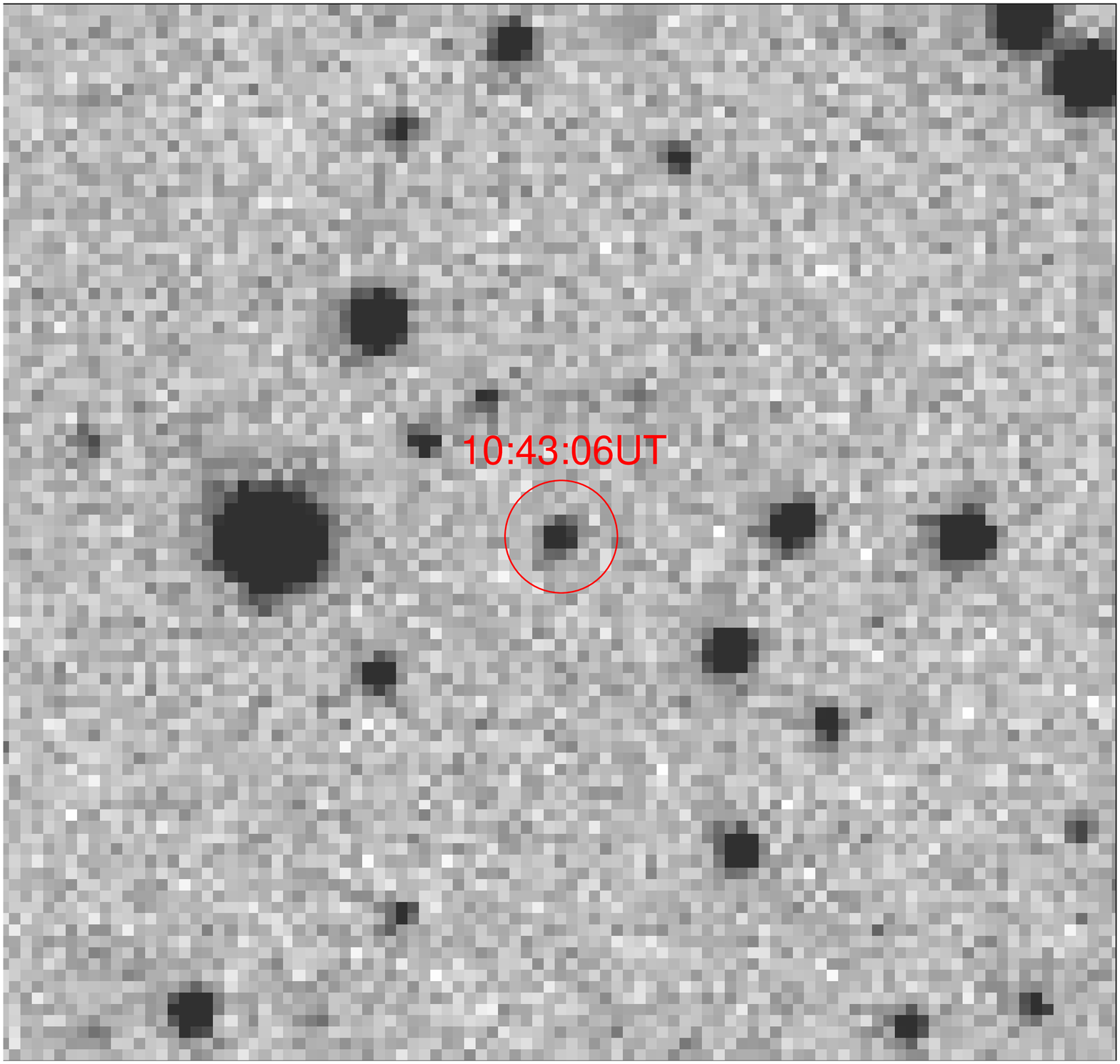}
   \includegraphics[width=0.3\textwidth]{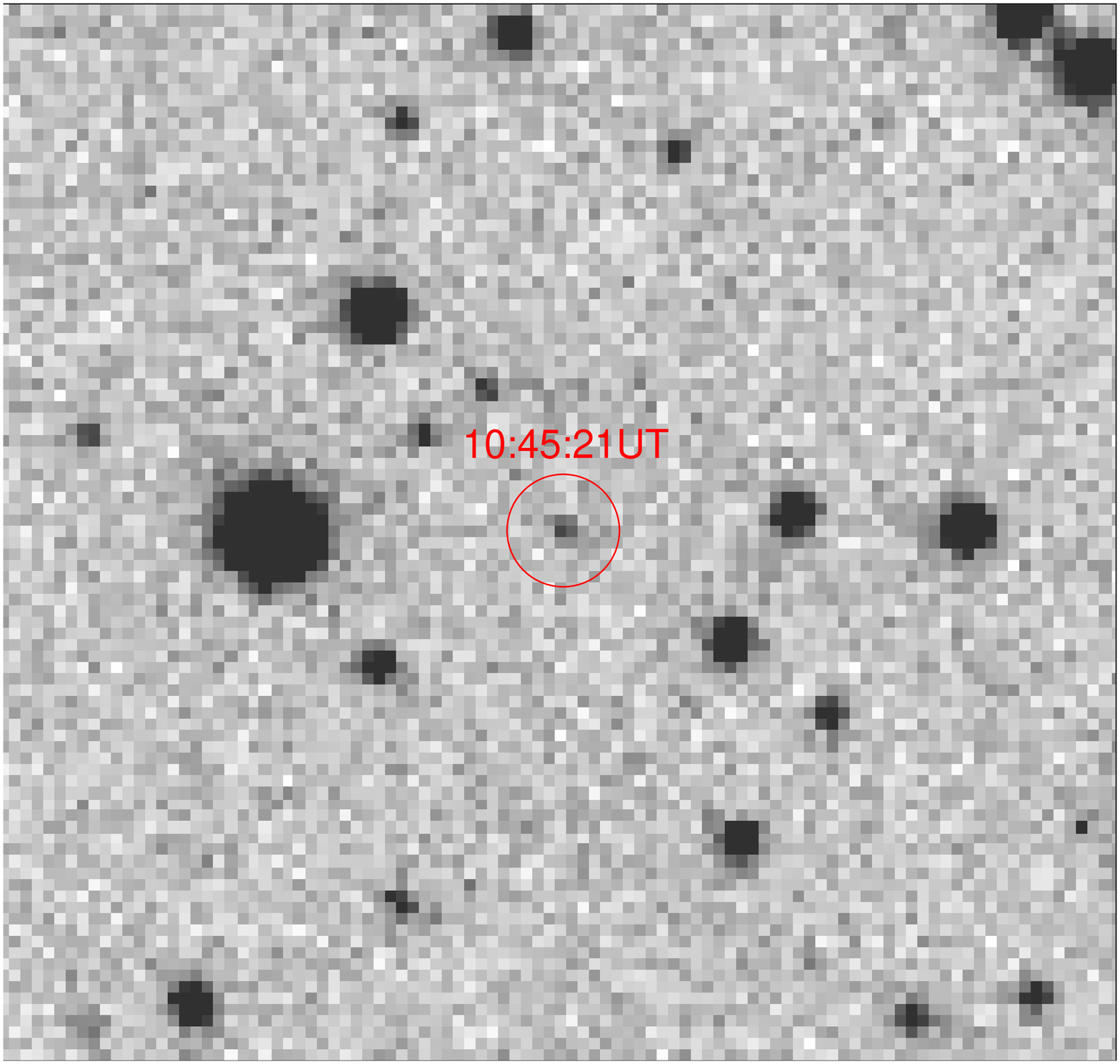}
 \caption{Finding Charts of GWAC\,181229A detected by GWAC. All these images were obtained by GWAC at the same night, and all the observation times are marked. 
 The left panel is the reference image that was obtained at about 38 minutes before the onset of the event. 
 The right two panels are the images taken after the onset.  
 The central source marked by the red circles is the object. There was a clear fainting during our observations. 
  }
 \label{findchart}
 \end{figure}

The detection magnitude was 13.5 mag in $R-$band measured by the real-time pipelines.
The coordinate of the new source measured from the GWAC images is R.A.=01:33:33.08, DEC=00:32:23.02 (J2000). 
The corresponding astrometric precise is about 2.0 arcsecond typically (1$\sigma$). 
This source was not detected in the reference image which was obtained by stacking 10  images taken at around 10:04:21 UT, 
i.e., about 38 min before the trigger time. 
The finding charts of the detection image and the reference image observed by GWAC are shown in Figure.\ref{findchart}. 
The candidate shows  stellar profile indicating that it is likely not originated from hot pixel, fast moving objects or ghosts  in GWAC system.
No any apparent moving was obtained by the pipeline for the transient. No any known minor planet or comet brighter than $V=20.0$mag was found in the 15.0 arcminute region around the transient\footnote{https://minorplanetcenter.net/cgi-bin/mpcheck.cgi?}. All these information indicates that the transient is a real astronomical event with a high level of confidence.

The on-line data processing showed that the transient fading by 0.9 mag can be seen in all the single exposures within a duration of 2.5 minutes after the first detection by GWAC. 
The detection limit of all these single exposures was $R\sim$15.0 mag at a significance level of 3$\sigma$.

We re-performed an off-line pipeline with a standard aperture photometry at the location of the transient and for several nearby bright reference stars by using the IRAF APPHOT package, including the corrections of bias, dark and flat-field in a standard manner.
After a differential photometry, the finally calibrated brightness of transient was obtained by using the SDSS catalogues through the Lupton (2005) transformation \footnote{http://www.sdss.org/dr6/algorithms/sdssUBVRITransform.html\#Lupton2005 (R = r - 0.2936*(r - i) - 0.1439;  sigma = 0.0072) }.

\begin{figure}[htbp]
 \centering
  \includegraphics[width=0.295\textwidth]{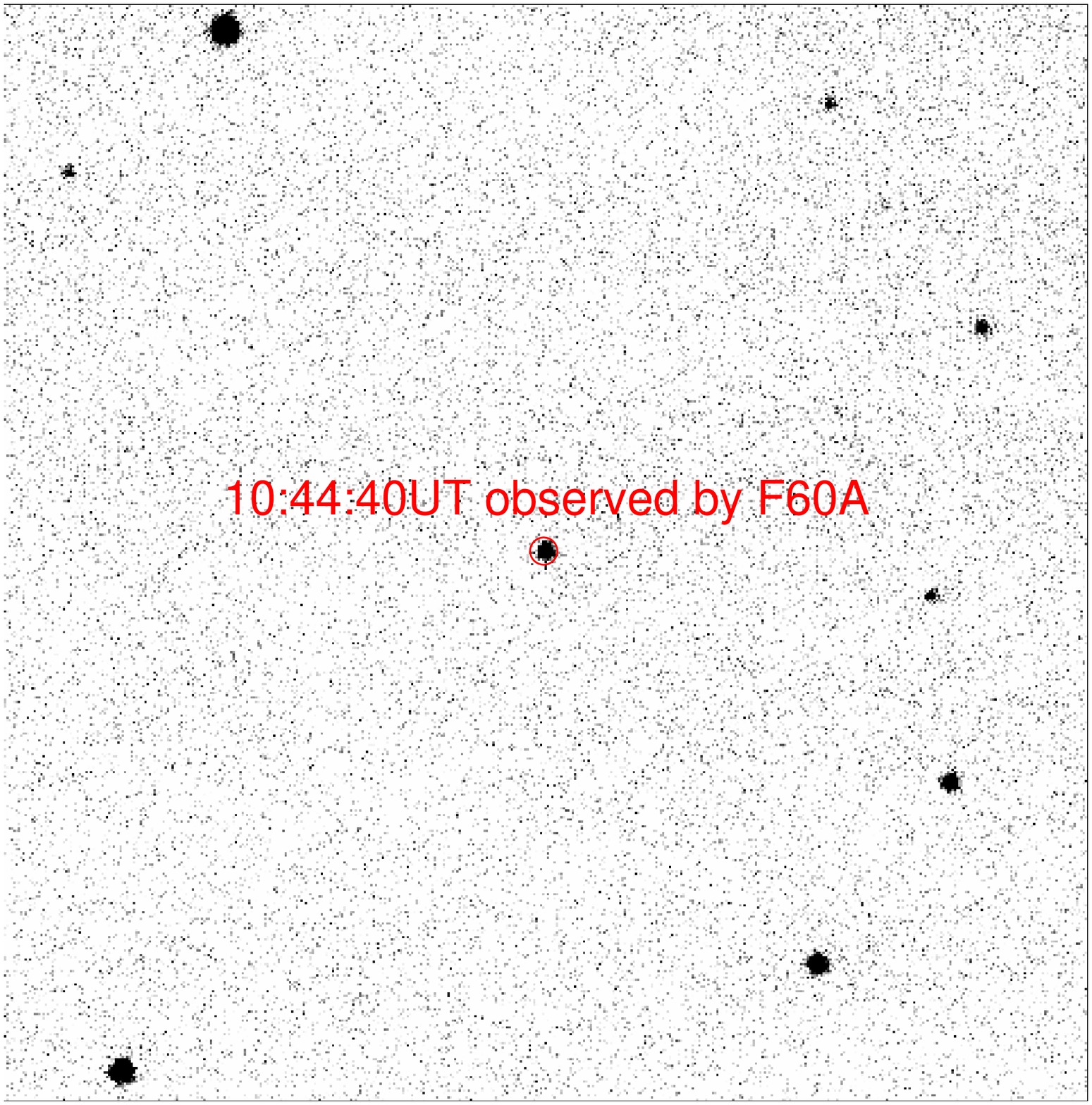}
   \includegraphics[width=0.295\textwidth]{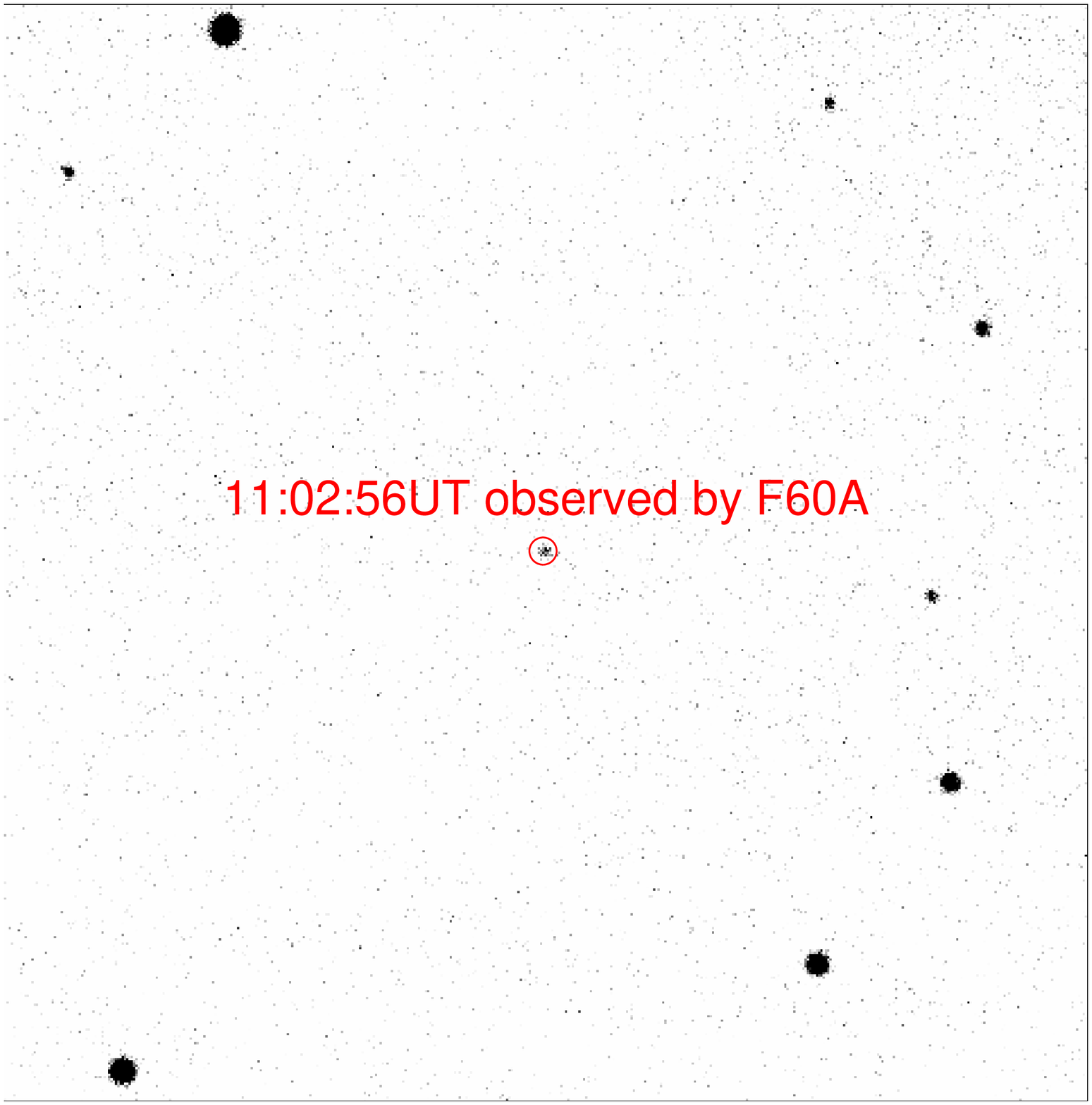}
    \includegraphics[width=0.3\textwidth]{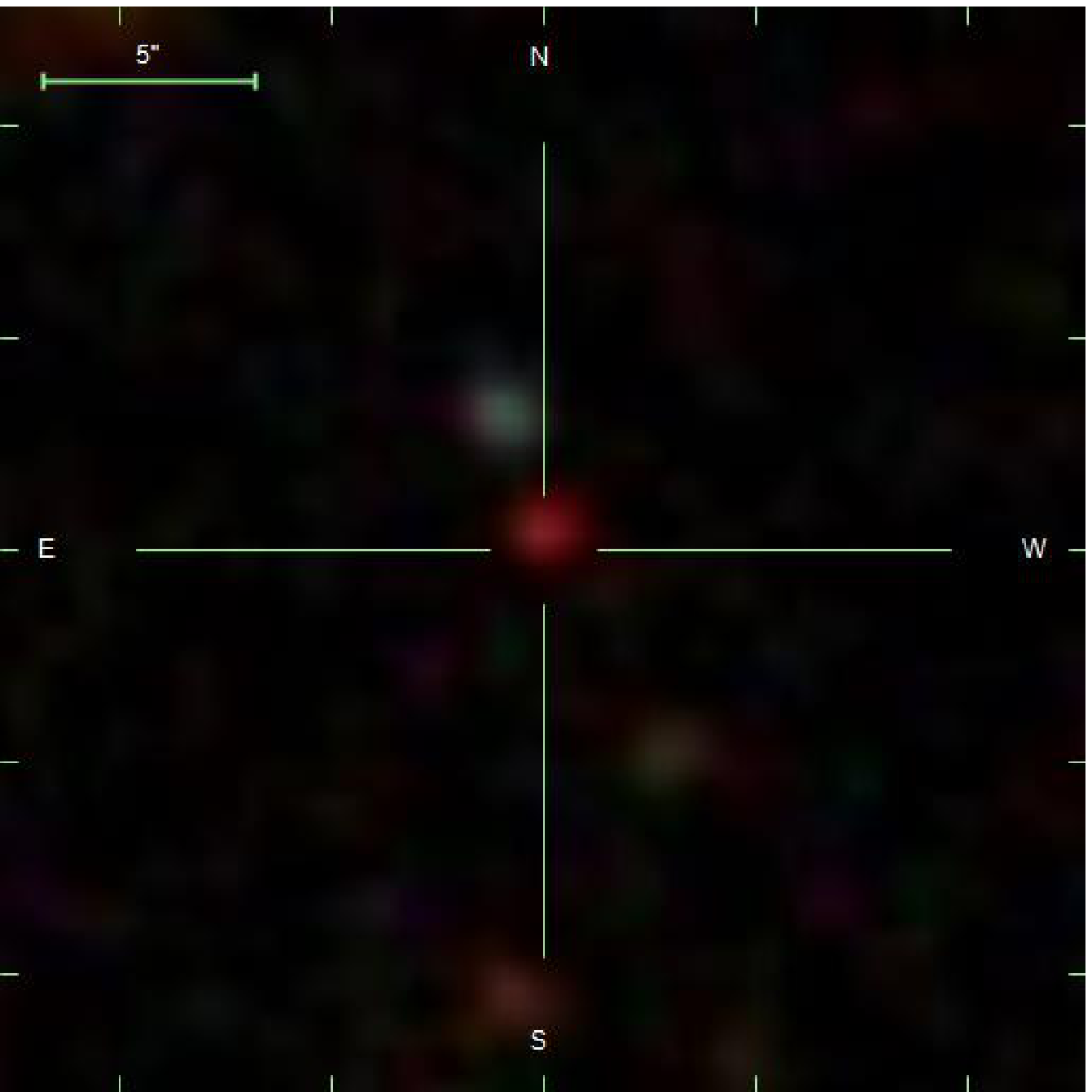}
 \caption{The left and middle panels are the finding Charts of GWAC\,181229A observed by F60A.  The field size is about 3.0 arcmin. The observation times in UTC on 2018, Dec. 29 are labeled in the images. The sources marked in the images are the object GWAC\,181229A. The brightness of this object clearly fades out during  our observations. The right panel is derived from SDSS DR13 survey for a comparison.  The central red and faint source with a magnitude of $r$=24.05 mag (Annis et al. 2014) is the counterpart of the flare. The celestial distance of the object from the position derived from F60A to the SDSS source is 0.695 arcsec.  The size of the right panels is different from the left and middle ones only for a clarity of display. }
 \label{findchart2}
 \end{figure}

\section{Follow-ups by Imaging and Spectroscopy}
\subsection{Photometries by F60A}
Upon the flare was triggered by the GWAC real-time pipeline, it was immediately followed-up by F60A\footnote{The diameter is 60cm, the f-ratio is 8.0. The detector equipped on the mount is Andor 2k*2k CCD. The pixel scale is 0.52 arcseconds. }
in standard Johnson-Cousins $R-$band via a dedicated real-time automatic transient validation system (RAVS, Xu et al. 2020) that is developed to confirm candidates triggered by GWAC and  to obtain an adaptive light-curve sampling for an identified target. 
With RAVS, the exposure time can be dynamically adjusted automatically based on the evolution of brightness of an object. 
For the case of GWAC\,181229A, the range of exposure time is from 30 sec to 150 sec.
The follow-up observations by F60A started at 2 minutes after the trigger, and stopped at the time
when the object was fainter than the 
detection limit of $\sim$19.0 mag, which corresponds to a total duration of about 120 min.

The raw images were reduced by following the standard routine in the IRAF\footnote{IRAF is distributed by the National Optical Astronomical Observatories, which are operated by the Association of Universities for Research in Astronomy, Inc., under cooperative agreement with the National Science Foundation.} package, including bias and flat-field corrections.
The correction of dark current was not made since the impact for the photometry can be negligible with the CCD cooling down to $-70$ deg.
After an aperture photometry, absolute photometric calibration was 
performed with several nearby comparison stars with the 
Lupton (2005) transformation from SDSS data Release 14 catalog to the Johnson-Cousins system\footnote{http://www.sdss.org/dr6/algorithms/sdssUBVRITransform.html\#Lupton2005}.

Figure 2 compares the Sloan Digital Sky Survey (SDSS) image centered at the target to the images obtained by F60A, 
in which there is a faint red counterpart within a distance of 0.697 arcseconds between the locations measured by F60A and reported by  SDSS Stripe 82 catalogue  (SDSS J013333.08+003223.7, Annis et al. 2014). 
Its brightness is $r=24.05\pm$0.15 mag (Annis et al. 2014) , which is taken as the quiescent brightness for the further analysis.

\subsection{Spectroscopic Observation}

One long-slit spectrum was obtained by the NAOC 2.16 m telescope (Fan et al. 2016) by using the Beijing Faint Object Spectrograph and Camera (BFOSC)\footnote{The BFOSC spectrograph is equipped with a back-illuminated E2V55-30 AIMO CCD. } via a ToO request. 
The start observation time for the spectrum was at 11:21:51.0 UT,
39 minutes after the trigger time. 
The exposure time was 30 minute. 
The coverage of the exposure time during the flare is shown with the yellow vertical area in Figure.\ref{fig:LC_C00090}.
With a slit width of 1.8 arcsec oriented in the south-north direction, the spectral resolution is $\sim$10$\AA$  when grating G4 was used, which results in a  wavelength  coverage  of  3850-8000$\AA$. The wavelength calibration was carried out with the iron-argon comparison lamps.
Standard procedures were adopted to reduce the two-dimensional spectra by using the IRAF package, including bias subtraction and flat-field correction. The extracted one-dimensional spectrum was then calibrated in wavelength and in flux by the corresponding comparison lamp and standard calibration stars.

\section {Results and Analysis}
In this section, we investigate the nature of the quiescent counterpart of GWAC\,181229A from multi-wavelength catalogs. 
The properties of the flare is then analyzed by modeling the light curve, which yields an estimation of the total energy emitted during the flare.

\subsection{The quiescent counterpart}

\begin{table}
\begin{center}
\caption{Properties of SDSSJ0133 (the quiescent counterpart of GWAC\,181229A) extracted from various surveys.}
\begin{tabular}{cc} 
\hline Parameter & Value \\
\hline & SDSS J013333.08+003223.7 (Annis et al. 2014) \\
\hline R.A. & 23.38779  \\
Decl. & 0.53991 \\
$u$ & $28.5450 \pm 2.1725$ \\
$g$ & $25.5569 \pm 0.4284$ \\
$r$ & $24.0556 \pm 0.1538$ \\
$i$ & $21.0491 \pm 0.0179$ \\
$z$ & $19.4138 \pm 0.0137$ \\
\hline & Pan-Starrs DR1 (108640233878278191, Chambers et al. 2016) \\
\hline R.A. & 23.387840550  \\
decl. & +00.539781430 \\
$i$ & $20.8993 \pm 0.0630$ \\
$z$ & $19.6418 \pm 0.0360$ \\
\hline & AllWISE Data Release (J013333.07+003222.9,  Cutri et al., 2013)  \\
\hline
R.A. & 23.38787  \\
Decl. & 0.53992 \\
$W 1$ & $15.366 \pm 0.049$ \\
$W 2$ & $15.517 \pm 0.152$ \\
\hline & UKIDSS-DR9 Large Area Survey \\
& (J013333.07+003223.7, Lawrence et al., 2012; Ahmed et al. 2019)  \\
\hline
$Y$ & $17.97 \pm 0.03$ \\
$J$ & $17.11 \pm 0.02$ \\
$H$ & $16.52 \pm 0.03$ \\
$K$ & $16.10 \pm 0.03 $ \\
Spectral type & M9 \\
Dis & 144.6 pc \\
\hline
\label{Survey}
\end{tabular}
\end{center}
\end{table}

In order to make a further investigation on the nature of this object, it is crucial to analyze the properties of the object in the quiescent state. We retrieved photometries from the Sloan Digital Sky Survey (SDSS: York et al. 2000),  Wide  field Infrared Survey Explorer (WISE; Wright et al. 2010), Pan-STARRS DR1 catalogue (PS1, Chambers et al. 2016) and other catalogues based on a coordinate cross-match through the VizieR Service\footnote{https://vizier.u-strasbg.fr/viz-bin/VizieR}.
Each catalog returns only one source, named as SDSSJ0133,  within our search radius of 2 arcsec. 
Parts of the queried parameters are shown in Table \ref{Survey}. 

At the beginning,
based on the color-magnitude transformations given in Lupton et al. (2005)\footnote{http://www.sdss3.org/dr8/algorithms/sdssUBVRITransform.php},
we estimate a quiescent brightness in $R-$band of 23.03 mag, 
which results in a flare magnitude as large as $\Delta R=9.5$ mag.
The derived quiescent flux is  $F_{R,q} = 1.4 \times 10^{-18}$  erg  cm$^{-2}$ s$^{-1}$ $\AA^{-1} $ 
by converting the quiescent magnitude above with 
the zero flux and the transformation for R band (Bessel et al., 1998).
Ahmed et al., (2019) reported that the quiescent counterpart is a spectral type of M9. 
Due to the faint brightness of this source, no parallax or other report about the distance 
 including the Gaia DR2 catalogue (Gaia Collaboration 2018).
With the corresponding SDSS $i-$ and $z-$ bands magnitudes, based on the relation of color ($i-z$) and
the absolute magnitude provided by Bochanski et al., (2020, 2012), 
an absolute magnitude of $M_{r}$ = 17.7 mag  for quiescent counterpart is derived. 
Consequently, a distance of 
$d\sim$155.8 pc can be calculated with the estimation of 
the absolute magnitude and the apparent magnitude above.
The reddening effect could be neglect for the above colors and the derived spectral type, 
since the extinction in the Galactic plane along the line of sight 
is not significant with E(B-V)=0.021\footnote{https://ned.ipac.caltech.edu/}. 
This distance  is roughly consistent with the value of 144.6 pc reported by Ahmed et al., (2019). In the following analysis, 
the mean value of the distance of 150 pc will be used for further analysis.

However, it is noted that a spectral type of M7 would be obtained if the estimation is based on the $i-z$ value provided by the PS1 catalogue.
The difference  in the derived spectral type 
is possibly caused by the difference between PanSTARRS and SDSS filters. 
The alternative possibility is that SDSSJ0133 is active with a low amplitude at the PS1 survey time.
Other clue for an activity is the blue WISE\footnote{Wide-field Infrared sky Explore}
 infrared color of $\sim -0.15$ with $W1(15.366\pm0.049)$ and $W2 (15.517\pm0.152)$ (Cutri et al., 2013), 
which is slightly bluer than the expectation ($W1-W2\sim0.2$ ) made from the empirical relationships for ultracool dwarfs reported in Schmidt et al. (2015).

According to the relation between metallicity and color of late type stars (Equation.3 in West et al. 2011), the metallicity-dependent 
parameter $\zeta$ is estimated to be 0.859,  which is slightly larger than the criterion of the classification of the subdwarf 
($\zeta<0.825$, L\'epine et al. 2007).

\subsection{The flare}

\begin{figure}[htbp]
 \centering
  \includegraphics[width=0.8\textwidth]{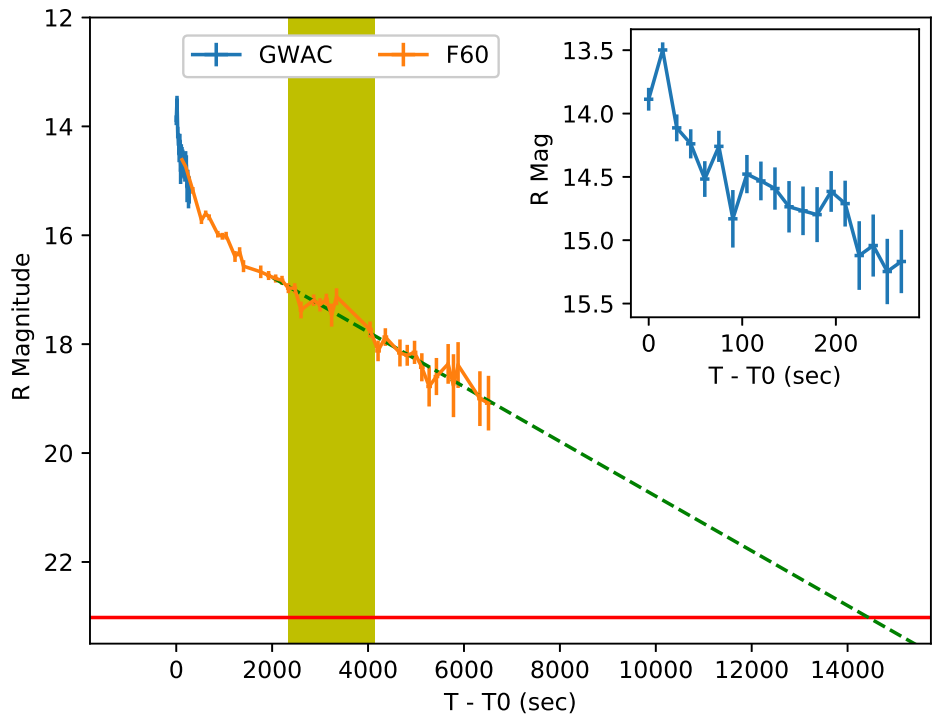}
 
 \caption{$R$-band light curve of GWAC\,181229A observed by GWAC and F60A. 
 The first detection occurs at $T_0=2458481.946482 $ day. The red line shows the quiescent brightness of this source with the magnitude of $R=23.03$ transformed from the SDSS $r$ and $i$ photometries. The green dash line presents the fitting result within the time interval of  [2000 sec, 7000 sec], and gives a prediction of the time for the end of the flare.  
 The inset panel shows the photometries obtained by GWAC around the peak time for more clarity.  
 The yellow vertical area  in the time interval of [2340 sec, 4140 sec] is for the exposure time ( 30 minutes ) of the spectrum observed by Xinglong 2.16m telescope.} 
 \label{fig:LC_C00090}
 \end{figure}

Figure \ref{fig:LC_C00090} shows the optical light curve of GWAC\,181229A,  in which the data taken by GWAC and by F60A is shown by 
blue and red points, respectively.
The horizontal red line marks the brightness level of the quiescent counterpart. The zoom panel at the upper right corner 
shows the GWAC data around the peak time. 
Before the first detection, the long-term monitors give an upper limit of 15.3 mag in $R$ band.  
At late phase, there are some fluctuations at low confidence since the signal-to-noise ratio decreases with time.  
The vertical errorbars are measurement-by-measurement estimates of the  photon statistical  error  including instrumental  characteristics. 
The horizontal errorbars correspond to 10 second exposure duration.

With a cadence of 15 seconds, the first detection of GWAC\,181229A shows that the brightness of the object was 13.9 mag in $R$ band, and  
the second one reaches the peak  with a brightness of 13.5 mag.
The brightness then falls to less than half the maximum only in two images with 30 seconds.
The total duration of the flare from the onset to the quiescent flux level is estimated to be about 14,465 seconds
by assuming that the brightness fades with a constant slope determined by fitting the late data as shown in Figure.\ref{fig:LC_C00090}. 

\subsection{Model the light curve} 

In order to have a more precise description of  the morphology of the flare of GWAC\,181229A, 
we fit the light curve for the decay phase  after the peak time by following the procedure of Davenport et al. (2014) (D14) ,  who tried to build a template from the 
single peak flares detected in active flare star GJ\,1243.  
Their procedure is as follows. For each flare, the flux and time after the onset are normalized to the quiescent level and 
the full time width at half the maximum flux ($t_{1/2}$), respectively.
The key parameter $t_{1/2}$  can be obtained by 1) fitting the light curve as a free parameter; 2) estimating in advance 
if the sampling of the light curve around the peak is dense enough.
The decaying light curve is described by a sum of two exponential curves as presented by Eq.4 in D14, standing for the two components: 
the impulsive decay phase and the gradual decay phase.

 For the case of GWAC\,181229, 
 the uncertainty of peak time is less than 7.5 seconds due to the GWAC's short cadence of 15 seconds. 
 By assuming that the peak magnitude we detected is the real peak brightness of the flare,
 the amplitude of $\Delta R\sim9.5$ mag corresponds to the relative flux of $F_{\text{amp}}=6500$
 which will be fixed during the analysis in our work. 
  We here model the rising and the decaying phase separately as follows.

 \subsubsection{Rising phase}
In the template of D14, the rising phase is fitted with a fourth-order polynomial. However,
 for the case of GWAC\, 181229A, before the peak time, most of the observation data are upper limits except for one real detection. 
 The behavior could not be well constrained with a fourth-order polynomial as the template of D14. 
 Here we have only to describe the rising phase of the flare briefly by assuming that this part follows a linear curve for  few detections. 
 \begin{equation} 
 F_{\text {decay}}/F_{\text{amp}}= k_0 + k_1 t         
  \end{equation}
  where $F_{\text{decay}}$ is the relative flux and $F_{\text{amp}}$ the peak relative flux that is fixed to be 6500.
 The values of $k_0$ and $k_1$ are calculated to be 0.69 and 0.02, respectively.  
 The uncertainties of two parameters can not be well estimated since there are only one positive detection before the peak. 
 The uncertainties of these values are about 10\% if only the  precise of photometry measurements are taken into account. 
  With this model, the onset time for the flare is about 35 seconds before the first detection, or 50 seconds before the peak time.

\subsubsection{Decaying phase}

\begin{figure}[htbp]
 \centering
  \includegraphics[width=0.4\textwidth]{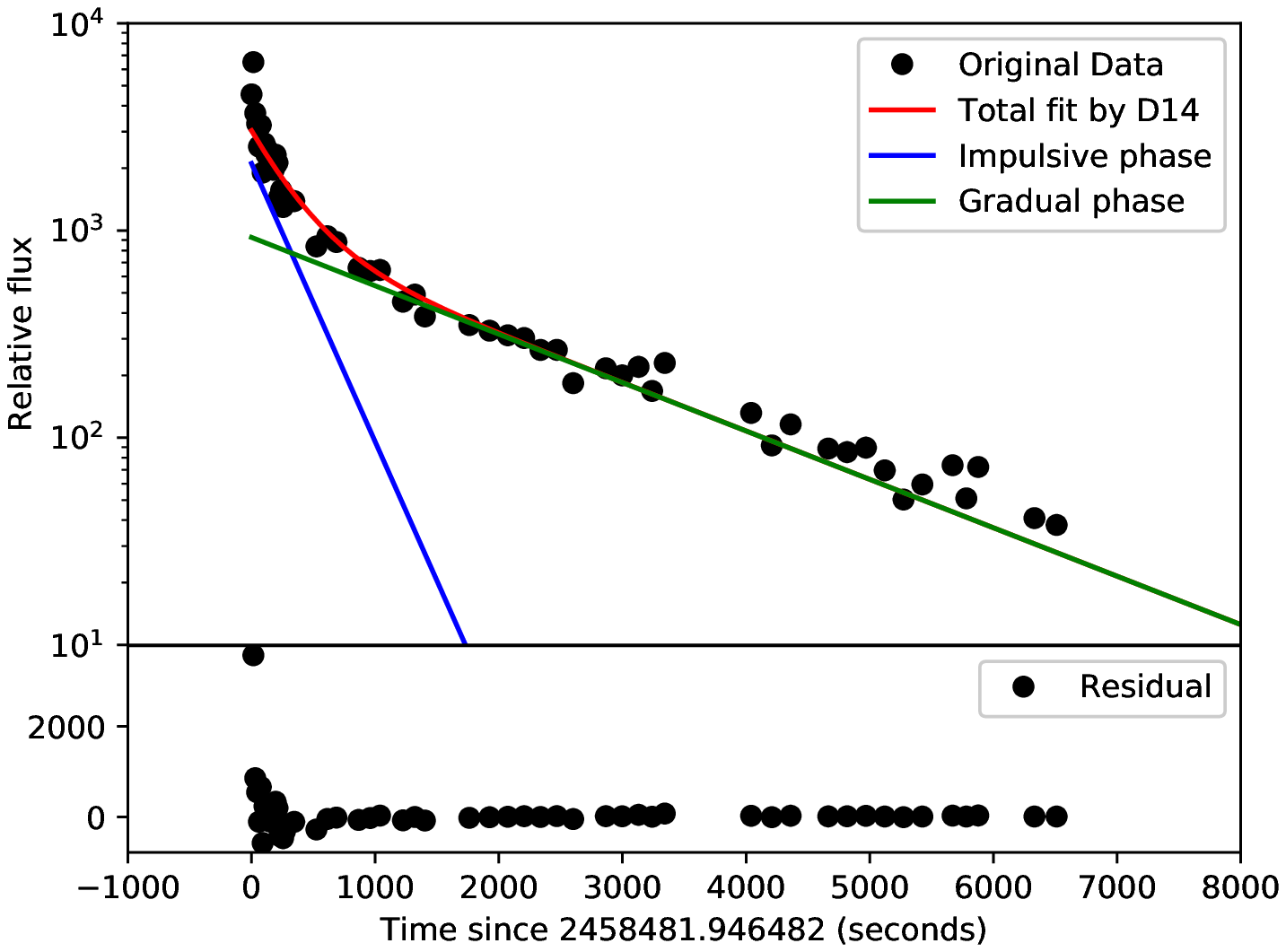} 
   \includegraphics[width=0.4\textwidth]{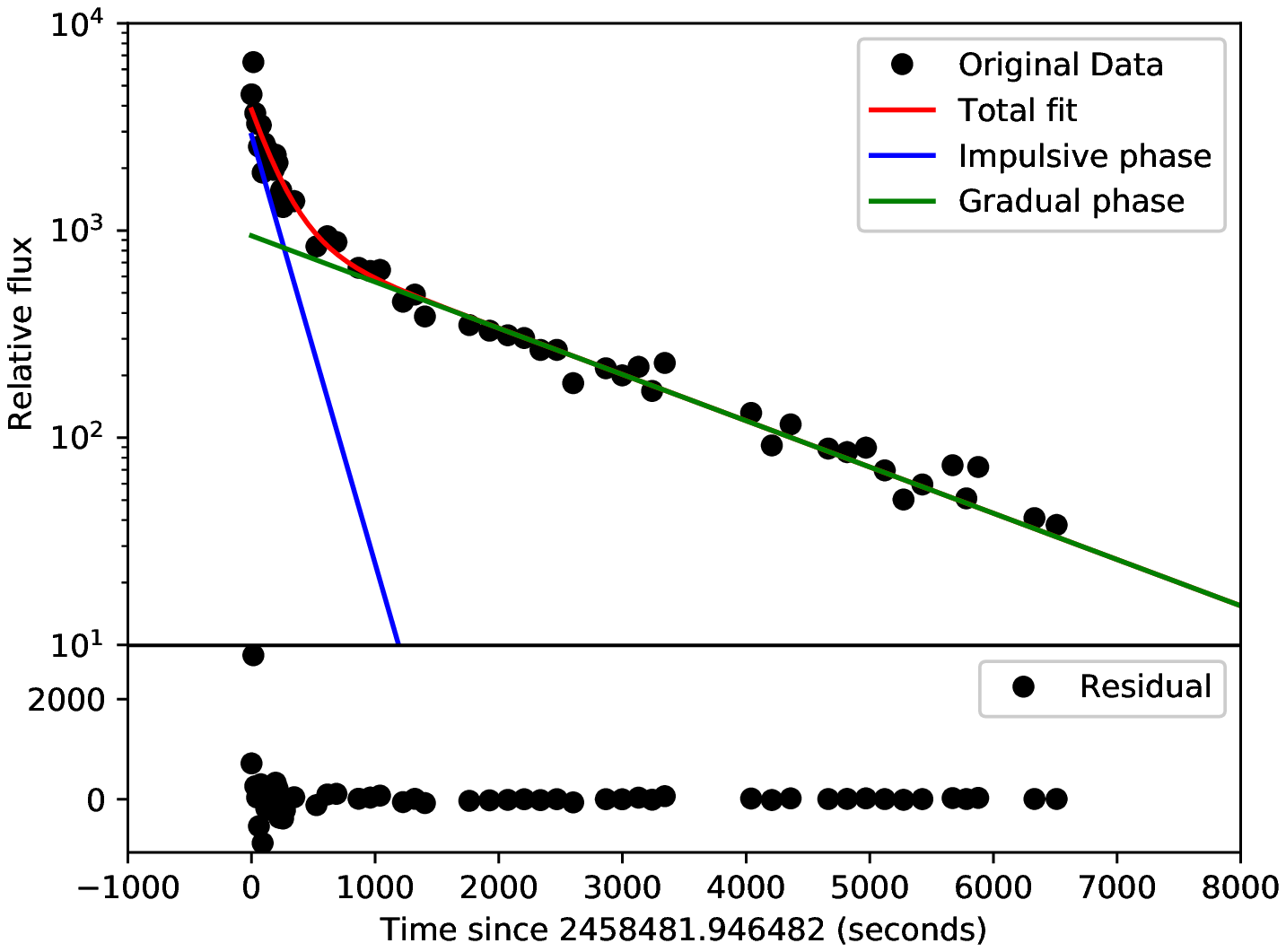} 
 \caption{\it Left panel: \rm Black data is the optical light curve of GWAC\,181229A observed by GWAC and F60A. Y-axis is the relative flux,
 and X-axis the time since $T_0=2458481.946482 $ day when the flare was first detected by GWAC. The red line shows the best fitted model
 described by a sum of two exponential laws. 
 The blue and the green lines present the impulsive and gradual components, respectively. 
 The left low panel gives the residual  for each data. 
 \it Right panel: \rm The same as the left one, but for the fitting in which the parameters is set to be free, except for the peak flux and the time scale unit.  It is clear that the peak brightness deviates from the expectations of the two fittings, indicating that the data near the peak time are originating from an additional more steeper component.} 
 \label{fig:D14fit}
 \end{figure}

After the modeling of the rising phase, we started from examining whether the D14 model can fit the observed data in the decaying phase.
In D14, a sum of two exponential laws as shown in the Equation \ref{func2} was adopted to describe the light curve. 
\begin{equation} 
F_{\text {decay}}/F_{\text{amp}}= k_1e^{-\alpha_1t/t_{1/2}} + k_2e^{-\alpha_2 t/t_{1/2} } 
\label{func2}       
\end{equation}
where $k_1=0.6890\pm0.0008$, $k_2=0.3030\pm0.0009$, $\alpha_1=1.600\pm0.003$, and $\alpha_2=0.2783\pm0.0007$ as given in D14 are fixed 
in the subsequent modeling. By setting the peak flux ($F_{\text{amp}}$) and the time scale $t_{1/2}$ as free parameters, the best fitting returns 
$F_{\text{amp}}=3059\pm63.6$ and $t_{1/2}=517.4\pm12.0$ seconds.  
The reduced $\chi^2/dof=3.63$ with a degree of freedom of 54. The large  $\chi^2$ indicates that the template of D14
does not provide a good fit to the data, especially near the peak time as shown in the left panel in Figure \ref{fig:D14fit}.
In fact,  by checking the light curve by eyes, the real $t_{1/2}$ should be around 30 seconds due to the sharp curve around the peak.

To improve the fitting, we set the parameters in Equation \ref{func2} to be free except for the  $F_{\text{amp}}=6500$, $t_{1/2}=1$.
The modeled values are tabulated in Table 3, and the reduced $\chi^2/dof=2.65$ with a degree of freedom of 52. 
The fitting results are shown in the right panel in Figure \ref{fig:D14fit}. In the upper panel of the figure, the total fitting result is displayed by the red line, and the two components with the blue and green lines, respectively. 
The time at which the  two components have equivalent contributions is 793 sec since the peak time. 
The lower panel shows the residual data that is obtained by a subtraction of the total fitting result from the observation data. 
The data near the peak time are still poorly reproduced, indicating that they might be from a new, more steeper component 
that is not included in the Equation \ref{func2}.

\begin{table}
\begin{center}
\caption{Parameters of the modeled decaying light curve of GWAC\,181229A. $\alpha_3$ is for the first impulsive decay phase. $\alpha_1$ and $\alpha_2$ stands for the gradual phase and shallow phase, respectively. }
\begin{tabular}{cccccc} 
\hline \hline 
$k_1$ & $k_2$ & $k_3$ & $\alpha_1$ & $\alpha_2$  & $\alpha_3$\\ 
\hline
\multicolumn{6}{c}{Two components model}\\
\hline 
$0.444\pm0.002$ & $0.145\pm0.007$ & \dotfill &  $0.005\pm0.001$ & $0.0005\pm0.0001$ & \dotfill \\
\hline
\multicolumn{6}{c}{Three components model}\\
\hline
$0.373\pm0.016$ & $0.128\pm0.008$  & $2.248\pm1.061$ &   $0.106\pm0.008$ &  $0.014\pm0.001$  & $2.946\pm0.895$ \\
\hline
\label{ModelingLC}
\end{tabular}
\end{center}
\end{table}

\begin{table}
\begin{center}
\caption{BIC for three models}
\begin{tabular}{cc} \\
\hline \hline 
model & BIC \\ 
\hline
D14 model  & 660.03\\
Two components model & 602.56\\
Three components model & 522.46\\
\hline
\label{BIC}
\end{tabular}
\end{center}
\end{table}

\begin{figure}[htbp]
 \centering
  \includegraphics[width=0.4\textwidth]{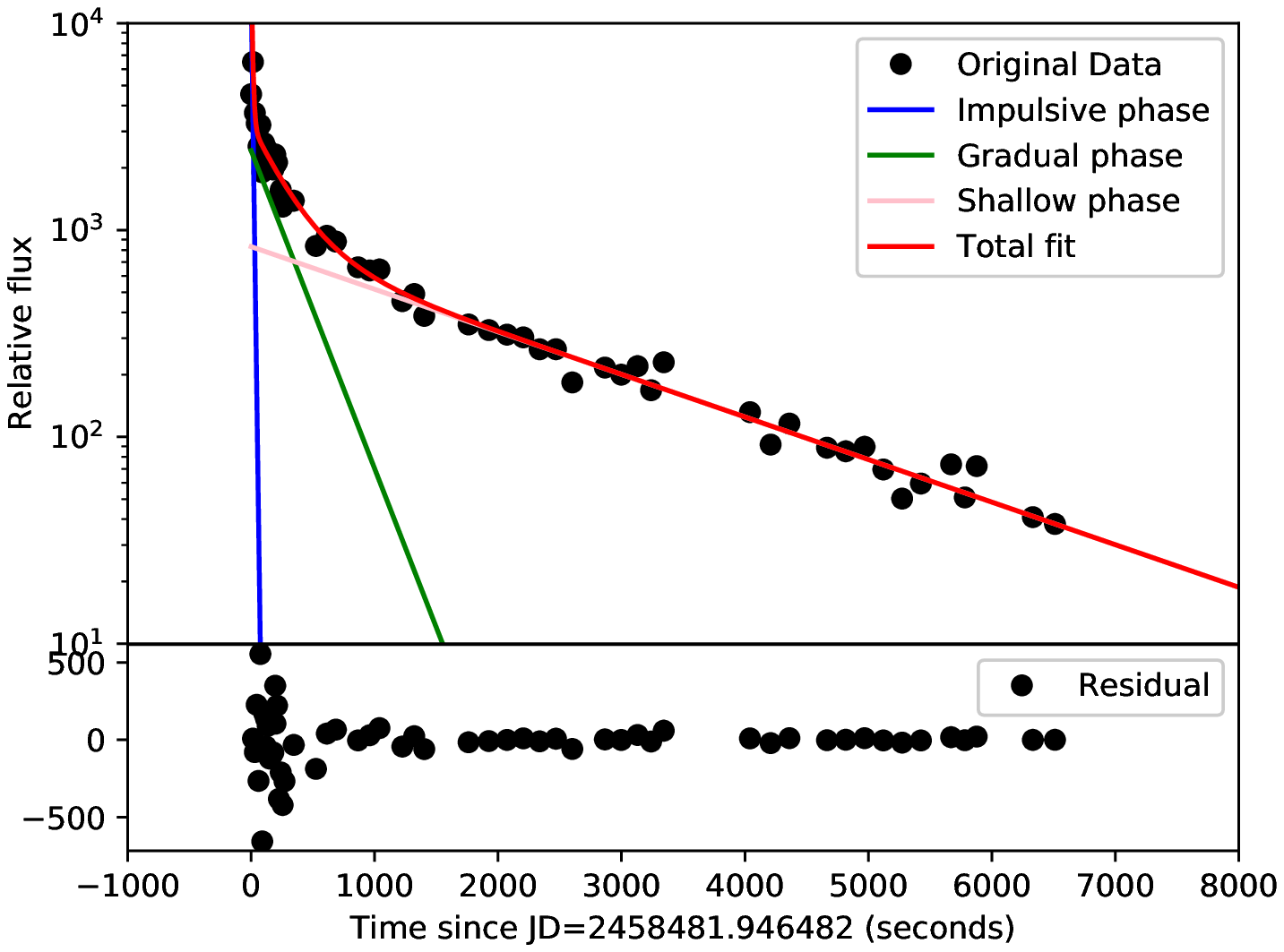} 
  \includegraphics[width=0.4\textwidth]{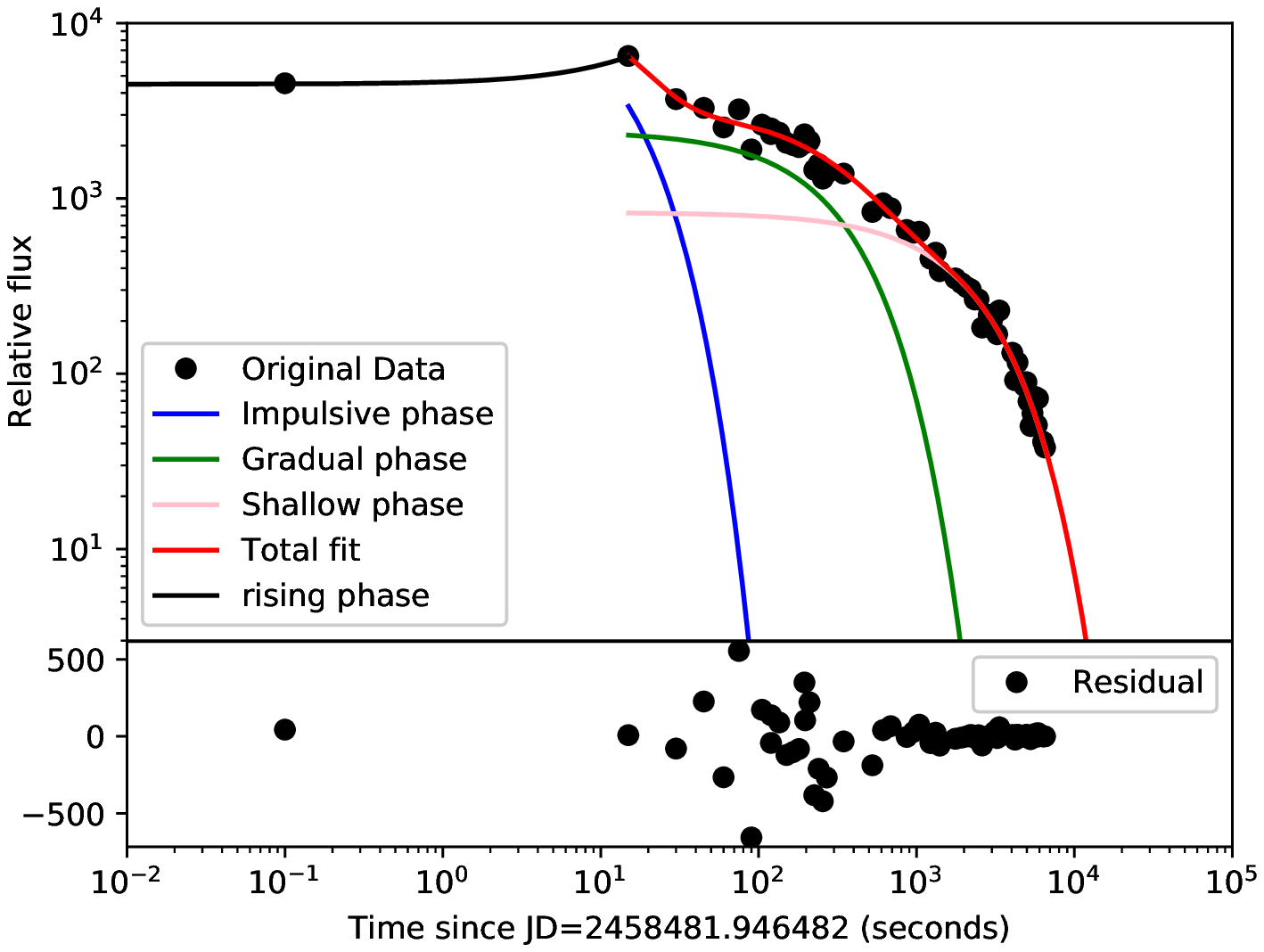} 
 \caption{Left panel:The same as Figure \ref{fig:D14fit}, but for a fitting with three components. Right panel: The same as the left but in the logarithmic scale for more clarity. In the right panel, the fitting result for rising phase with  black line is also displayed. The total fit in red line in two panels is only for the decay phase after the peak time.} 
 \label{fig:LC_C00090_fit_func3}
 \end{figure}

In order to reproduce the light curve around the peak, we then model the light curve in the decaying phase by adding an exponential component:
\begin{equation} 
F_{\text {decay}}/F_{\text{amp}}= k_1 e^{-\alpha_1t/t_{1/2}}+k_2e^{-\alpha_2t/t_{1/2}}+k_3e^{-\alpha_3t/t_{1/2} }     
 \label{func3}     
\end{equation}
A much better fitting with a reduced $\chi^2/dof=1.15$ with a degree of freedom of 50 can be learned from Figure \ref{fig:LC_C00090_fit_func3}. 
The modeled parameters are again listed in Table.\ref{ModelingLC}. 
This good fitting suggests that there are three components in the decay phase. 
After the peak time, there is a very sharp decay component. 
At the time around 75 seconds, the light curve transfers to the second gradual component. After about 1500 seconds, the third 
shallow decay is dominant until the end of the flare. 

A Bayesian information criterion (BIC) is used to test whether the three components model used in the fitting is required or  resulted from overfitting the data. 
The BIC values are 522.46, 660.03, 602.56 for three components model, D14 model, and two components model, respectively. 
All these BIC values are also summarized in Table.\ref{BIC}.
This result confirms that three components model is more reasonable for the data. 

Although some complex light curves has been observed (e.g., Kowalski et al. 2010), previous works presented that the morphology of flare light curves are typically divided into two phases: an impulsive phase and a gradual decay phase(e.g., Moffett 1974; Moffett \& Bopp 1976; Hawley \& Pettersen 1991; Davenport et al., 2014). 
However, for the case of GWAC\,181229A, three phases are needed to describe well the high-cadence light curves. The initial decay  is lasting to 20 sec after the first detection(5 sec after the peak time), which likely dominated by a brighter, hotter region that cools very shortly, and then a gradual decay phase from about 20 sec to 350 sec which corresponds to a cool region in which the radiation cools slowly. Finally, the event are moving to the last shallower decay phase  lasting from about 350 sec to the quiescent state.

\subsubsection{Ratio of decay indices}

We define the ratio of decay indices, donated by ${R_{ij}}=\alpha_{i} / \alpha_{j}$ ($i,j=1,2,3$),
to present how  fast the cooling speed changes from one phase to another, which is independent on the time unit scale $t_{1/2}$. 
For the case of GWAC\,181229A, they are deduced to be ${R_{31}}\sim27.74$  from the impulsive decay phase to the gradual phase, and ${R_{12}}\sim7.47$ from the gradual phase to the shallow decay phase, respectively. 
To make a comparison,
the value of ${R}$ from the template derived by D14 is $\alpha_{D1}/\alpha_{D2}=1.600/0.2783=5.749$. Such a difference might be attributed to the possible dependence on properties such as stellar effective temperature or magnetic field strength during the flares.

\subsection{Spectrum properties}
\begin{figure}[htbp]
 \centering
 \includegraphics[width=0.6\textwidth]{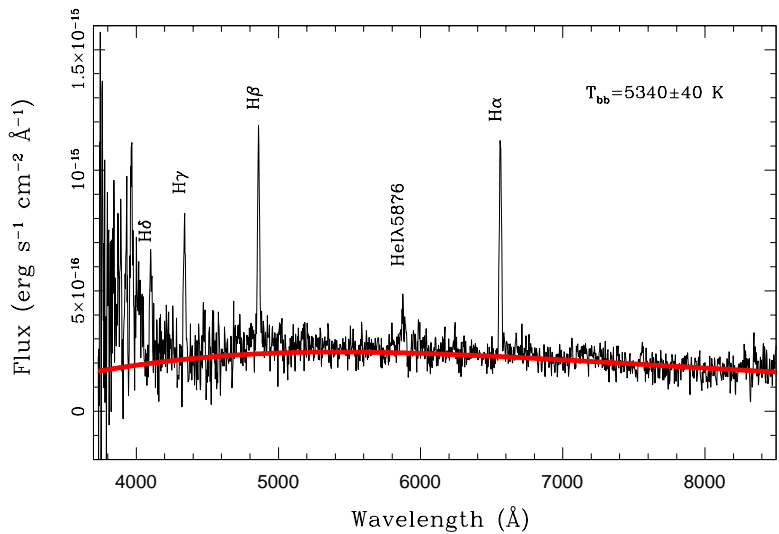}
 \caption{The spectrum obtained by  the 2.16m telescope at Xinglong observatory, China. A modeling of the underlying continuum by 
 a hot black body is shown by the heavy red line. }
 \label{spec}
 \end{figure}

Figure \ref{spec} shows the spectrum taken by the 2.16m telescope. A series of  strong emission lines such as $\mathrm{H\alpha}$, 
\ion{He}{1}$\lambda5876$,  $\mathrm{H\beta}$, $\mathrm{H\gamma}$ and  $\mathrm{H\delta}$ are marked on the spectrum.
The fluxes measured by a direct integration are presented in Table.\ref{spec_line}. 
After excluding the regions with the strong emission lines,
we modeled the underlying continuum by  a black body in the wavelength range 4000-8000\AA, which returns  a temperature of $T_{\mathrm{bb}} = 5340\pm40$K.

These emission lines are commonly detected during a dMe flare (e.g., Kowalski et al., 2013) 
and thought to be associated with chromospheric temperatures. 
By summarising the flux of these strong emission lines shown in Table.\ref{spec_line}, the total energy in the emission lines of $4.8\times10^{-14} \mathrm{ erg/s/cm^{2}}$ in our observation wavelength range could be derived. The total emission of $5.13\times10^{-13} \mathrm{ erg/s/cm^{2}}$ for the continuum emission within the wavelength range from 4000 to 8000 $\AA$  also be measured. 
The ratio of the energy in the emission lines and the underlaying continuum is about $\sim$9.3\% for GWAC\, 181229A, which is higher than the percentage ($\sim$4\%) in the impulsive phase (Hawley \& Pettersen 1991)  and is significantly smaller than the values (17\%-50\%) in the gradual decay phase reported in the literatures ( e.g., Hawley \& Pettersen 1991; Hawley et al. 2007).

Previous works in the literatures show that the temperature at gradual phase is lower than the values obtained at peak time (e.g., Fuhrmeister et al., 2008; Schmitt et al., 2008). 
Our measured temperature  of $\sim$5340 K  in the shallow decay phase is similar with the reported temperature of 5500-7000K in the decay phase of a flare event presented by Mochnacki \& Zirin (1980), but is slightly higher than the reported values in the decay phase (Fuhrmeister et a., 2008; Schmitt et al., 2008) where a blackbody temperatures of  3200-5600 K was given after measuring the continuum shape in  their red higher cadence spectra.

\begin{table}
\begin{center}
\caption{Emission line measurements of the spectrum of GWAC\,181229A displayed in the Figure \ref{spec}}
\begin{tabular}{cc} 
\hline \hline 
Line  &       Flux ($\mathrm{10^{-15}\,erg\,s^{-1}\,cm^{-2})}$\\
\hline 
$\mathrm{H\alpha}$ &        16.15 \\ 
\ion{He}{1}$\lambda$5876   &  2.79 \\
$\mathrm{H\beta}$      &   13.64 \\
$\mathrm{H\gamma}$       &   9.28  \\
$\mathrm{H\delta}$     &     6.50  \\
\hline
\label{spec_line}
\end{tabular}
\end{center}
\end{table}

\subsection{Energy budget}
The equivalent duration ($ED$) of a flare is defined to be the time needed to emit all the flare energy at a quiescent flux level  (e.g. Kowalski et al. 2013). 
By integrating the model of the light curve over the range of the light curve from the start to the end of the flare, 
the $ED$ is estimated to be $\sim2.584601\times10^{6}$ seconds, 
or 29.9125 days for GWAC\,181229A. 
Following the method of Kowalski et al., (2013),
the total energy $E_{R}$ in $R-$band can be calculated
with the equation $E_{R}=4\pi r^2\times F_{R,q}\times ED$,
where
the quiescent flux $F_{R,q} = 1.4 \times 10^{-18}$  erg  cm$^{-2}$ s$^{-1}$ $\AA^{-1} $ and the distance is r=150 pc, the energy $E_{R}$ is measured to be $ 1.54 \times 10^{34} $ ergs.\footnote{It is noticed that there is a caveat that this method  is based on a simple assumption that the flare spectrum is similar to the one 
in the quiescent state which is however not fully consistent with fact. The uncertainty for the estimated energy shall be within 8\% as a maximum value with the different blackbody spectrum shape from T=10 000K to T=2300K. }

To estimate the bolometric energy, one have to get the knowledge the effective temperature. 
In this work, our spectrum during the decay phase gives a temperature of $5430\pm40$ K by a blackbody spectrum fit.
On the other hand, the temperature during the peak time for a dMe flare could be as high as $T_{eff}=10^{4} K$ (e.g., Kowalski et al. 2013).
More evidences indicate that the temperature shall be evolving during the flare from peak time to the gradual decay phase (e.g., Hawley \& Pettersen 1991; Hawley \& Fisher 1992). 
Here for simplicity,  the bolometric energy will be estimated based on two effective temperatures, 
one is  $T_{eff}=10^{4} K$  and  
the other is $T_{eff}=5340 K$. 
By integrating the spectrum of a blackbody shape with effective temperatures shown above with the wavelength range from  1 nm to 3000 nm, and calibrated the energy with R band flux,
the bolometric energy $E_{bol}$ of $9.25\times 10^{34} $ ergs and $5.56\times 10^{34}$ ergs for 
$T_{eff}=10^{4} K$ and $T_{eff}=5340\pm40 K$ could be obtained, respectively. 
With the same method, the $U$-band energy of the flare is $E_U\sim1.5\times10^{34}$ ergs and $E_U\sim3.6\times10^{33}$ ergs for the two temperatures, respectively. 
Such a large amount of energy makes this flare to be comparable to the flare event  SDSSJ0221 ($E_U=(3.2-5.5)\times10^{34}$ ergs) reported by Schmidt et al. (2016)  and CZ Cnc reported by Schaefer (1990),  and to be one of the largest energy events from ultracool dwarfs.

\subsection{Continuum emission in $R$-band}

The flare emission at optical and $\text{UV}$ wavelengths are believed to be contributed by two major components. 
The dominated one is a hot blackbody emission (continuum emission) with a template of about $T\sim10,000$K (e.g., Hawley \& Fisher 1992)  that is considered to be produced at the bottom in the stellar atmosphere near the foot points of the magnetic field loops. 
The second component is the atomic emission lines (e.g., Fuhrmeister et al. 2010) and hydrogen Balmer continuum (Kunkel 1970). 
The proportion of the two contributors changes with the evolution of the flare. 
Near the peak time, the continuum emission could contribute more than 90\% emission ( Hawley \& Pettersen 1991) of the total energy of the flare.   
In the gradual phase, the fraction of the continuum can drop to 69\%( Hawley \& Pettersen 1991) or even down to 0\% (Hawley et al., 2003).

The filling factor $X_{\mathrm{fill}}$ is the fraction of the area of the projected visible stellar disk that emits flare continuum emission, which allows us to understand what type of heating distribution is responsible for the observed light curve (Kowalski et al. 2013). 
Following the method of Hawley et al. (2003), $X_{\mathrm{fill}}$ in the impulsive and gradual phase can be deduced from
\begin{equation}
F_{\lambda}=X_{\mathrm{fill}}\frac{R^{2}}{d^{2}} \pi B_{\lambda}(T)   
\label{Fillingfactor}  
\end{equation}
where $R$ is the stellar radius, $d$ the distance, and $T$ the characteristic temperature of the blackbody emission. 
$F_{\lambda}$ is the flare flux observed at Earth at wavelength $\lambda$, which can be measured from the optical spectrum 
within a range of wavelength free of emission lines. 

For the case of GWAC\,181229A, only one spectra was obtained at about 54 min after the event (mid time of the exposure as presented in Figure \ref{spec}).  The continuum flux level is measured to be $1.8\times10^{-16} \mathrm{ erg\ cm^{-2}\ s^{-1}\ \AA^{-1}}$
within the wavelength range of 6800-7200\AA. There is no any apparent emission lines within this wavelength range.
Adopting  $R=$ $0.1R_{\odot}$ for a typical radius of a M9 brown dwarf (Baraffe et al., 2015), 
$d=150$ pc, and a blackbody temperature of $T_{\mathrm{bb}}=5340$K yields a $X_{\mathrm{fill}}\sim$19.3\%
for the decay phase, by assuming that  all the emission measured within the wavelength range is produced by the blackbody emission.

Although there was no spectra obtained near the peak time, the temperature and the corresponding filling factor  $X_{\mathrm{fill}}$ can be estimated as follows. 
Assuming 95\% observed peak emission are contributed by continuum emission, 
a critical temperature $T_{\mathrm{c}}=10,000$K of a blackbody emission is deduced which corresponds to  a filling factor of 100\% of the surface of the object, indicating that the temperature of the blackbody emission near the peak time is much higher than the $T_c$. 
Further calculations are made with $T=16,000$K, $T=20,000$K, $T=30,000$K and $T=35,000$K to estimate $X_{\mathrm{fill}}$,
which results in a $X_{\mathrm{fill}}$ of 36\%, 24\%, 13\% and 10\%, respectively. 
We noted that  Kowalski et al. (2013) reported that
the  temperatures of the blackbody body is from $T=9800$ to 14100 K  for the peak of the flares of the mid-M dwarf.
If it is true for the later-M dwarf in GWAC\,181229A, the value of $X_{fill}$  is at the level of  $\sim 30\%$ at the peak time.

The maximum magnetic field strength $B_{z}^{max}$ associated with the super flare observed on GWAC\,181229A 
could be estimated with the scaling relation in Aulanier et al. (2013)  and Paudel et al., (2018) by assuming that the flare on GWAC\,181229A is similar with the solar flares.
\begin{equation} 
E_{bol}=0.5 \times 10^{32}\left(\frac{B_{z}^{\max }}{1000 \mathrm{G}}\right)^{2}\left(\frac{L^{\text {bipole }}}{50 \mathrm{Mm}}\right)^{3} erg
\label{MagneticStrength}  
\end{equation}
where $E_{bol}$ is the bolometric flare energy, and $L^{bipole}$ is the linear separation between bipoles. 
Since the filling factor X$_{\mathrm{fill}}$ is at the level of 30\% at the early phase, we could take $L^{bipole}$
as $\pi R$ as the maximum distance between a pair of magnetic poles on the surface of GWAC 181229A. 
With these parameters, a strong magnetic field of (3.6-4.7)kG is deduced. 
This strong magnetic strength is at the level of  the saturated value of 3-4 kG (Reiners et al., 2009), and slightly smaller than  the reported values of 7.0 kG for WX Ursae Majories  (Shulyak et al., 2017) and
5 kG for an M8.5 brown dwarf LSR J1835+3259 (Berdugina et al., 2017). 

\section{Summary}

In this paper, we report a giant stellar flare GWAC\,181229A detected by GWAC  with a survey cadence of 15 seconds. The peak brightness is measured to be $R=13.5$ mag. The counterpart of GWAC\,181229A is a M9 star with a brightness of $r$=24.0 (or $R$=23.03 mag), yielding an amplitude of 9.5 mag in $R$-band.  The total energy in $R$-band and the bolometric energy are estimated to be $1.5\times10^{34}$ erg, and $(5.56-9.25)\times10^{34}$ erg, respectively. 
The magnetic strength B is deduced to be (3.6-4.7)kG. 
Such huge energy budget places the flare  to be one of largest energy events for ultracool stars. 
A very fast follow-up observation in imaging  was carried out by F60A via RAVS with a delay of 2 min since the trigger time.
At 39 min after the trigger, a  low-resolution spectrum was started to be taken by the 2.16m optical telescope at Xinglong observatory, China. 

The flare promptly rises from the quiescent flux level to the peak time in about 50 sec, and then returns to a decay modeled by a combination of   three components which is required to properly reproduce the 
decaying light curve.  
Based on a fitting of the continuum emission in the spectrum by a blackbody, 
an effective temperature of $T=5340\pm40$ K.
The filling factor  is derived to be 19.3\% for the flare in the later gradual phase,
while it is 36\% at the peak if a temperature of $T=16,000 $K is adopted.

Thanks to the large field-of-view and the high survey cadence, GWAC is well-suited for the detection of white-light flares. Actually, 
we have hitherto detected more than $\sim130$ white-light flares with an amplitude more than 0.8 mag. 
More GWAC units are planed to work in the next two years, aiming to increase the detection rate of high amplitude stellar flare by monitoring more than 5000 square degrees simultaneously (Wei et al. 2016). This is essential for not only improving our understanding of the flares of late-type stars themselves, but also revealing the life-threatening on extrasolar planets by the largest flares. 
 
\section{Acknowledgement}
The authors thank the anonymous referee for a careful review and helpful suggestions that improved the manuscript. 
This study is supported from the National K\&D
Program of China (grant No. 2020YFE0202100)
 and the  National Natural Science Foundation of China (Grant No.
11533003, 11973055, U1831207). This work is supported by the Strategic Pioneer Program on Space Science, 
Chinese Academy of Sciences, grant Nos. XDA15052600 \& XDA15016500, 
 and  by the Strategic Priority Research Program of the Chinese Academy of Sciences, Grant No.XDB23040000. 
YGY is supported by the National Natural Science Foundation of China under grants 11873003. JW is supported by the National
Natural Science Foundation of China under grants 11473036 and 11273027.
We acknowledge the support of the staff of the Xinglong 2.16m telescope. This work was partially supported by the Open Project Program of the Key Laboratory of Optical Astronomy, National Astronomical Observatories, Chinese Academy of Sciences.
This work made use of data supplied by the UK Swift Science Data Centre at the University of Leicester.
This work has made use of data from the European Space Agency (ESA) mission
{\it Gaia} (\url{https://www.cosmos.esa.int/gaia}), processed by the {\it Gaia}
Data Processing and Analysis Consortium (DPAC,
\url{https://www.cosmos.esa.int/web/gaia/dpac/consortium}). Funding for the DPAC
has been provided by national institutions, in particular the institutions
participating in the {\it Gaia} Multilateral Agreement.
This research has made use of the VizieR catalogue access tool, CDS, Strasbourg, France (DOI: 10.26093/cds/vizier). The original description of the VizieR service was published in A\&AS 143, 23

\end{document}